\documentstyle[twoside,fleqn,espcrc2]{article}

\newcommand{\nn}{\nonumber}

\newcommand{\be}{\begin{equation}}
\newcommand{\ee}{\end{equation}}
\newcommand{\ben}{\begin{displaymath}}
\newcommand{\een}{\end{displaymath}}
\newcommand{\baa}{\begin{eqnarray}}
\newcommand{\eaa}{\end{eqnarray}}

\def\L{{\cal L}}
\def\T{\Theta}
\def\O{\Omega}
\def\c{\cite}

\def\g{\gamma}

\def\l{\lambda}

\def\bi{\bibitem}

\def\i{\infty}
\def\E{{\cal E}}

\def\x{\xi}
\def\xb{{\bar{\xi}}}
\def\be{\begin{equation}}
\def\ee{\end{equation}}
\def\la{\label}
\def\c{\cite}
\def\f{\frac}
\def\Eb{\bar{\cal E}}
\def\ba{\begin{array}}
\def\ea{\end{array}}

\def\Ref#1{(\ref{#1})}
\def\R{{\bf R}}

\def\nn{\nonumber}

\def\p{\partial}
\def\ra{\rightarrow}
\def\la{\label}
\def\Dt{\tilde{D}}
\def\Lh{\hat{\L}}

\title{Some remarks on finite-gap solutions of the Ernst equation}

\author{D.\,Korotkin\thanks{On leave of absence from Steklov
Mathematical Institute, Fontanka, 27, St.Petersburg, 191011 Russia.
\newline
e-mail: korotkin@x4u2.desy.de}
\thanks{This work was supported by DFG contract Ni 290/5-1}
\bigskip\\II. Institut f\"ur Theoretische
Physik, Universit\"at Hamburg, Luruper Chaussee 149, D-22761 Hamburg,
Germany}
\begin{document}
\begin{abstract}
It is explicitly shown  that
the class of algebro-geometrical 
(finite-gap)
solutions  of the Ernst equation constructed several
years ago in \c{K}  contains the   solutions
recently constructed by R.Meinel and G.Neugebauer \c{MN} as a subset.

\end{abstract}

\maketitle

\section{Algebro-geometrical solutions of Ernst equation}
The Ernst equations which arises from certain dimensional reduction of $4D$
Einstein's equations has the following form:
\be
(\E +\Eb)\Delta\E=
2(\E_x^2+ \E_\rho^2)
\la{ee}\ee
where $\E(x,\rho)$ is a complex-valued Ernst potential and 
\ben
\Delta = \p_x^2 +\f{1}{\rho}\p_\rho +\p_x^2
\een
is a cylindrical Laplacian operator.
For $\E\in\R$ Ernst equation reduces to the
classical Eulers-Darboux equation
\be
\Delta\log\E =0
\la{static}
\ee
corresponding to static space-times.
Denote $x+i\rho$ by $\x$ and
consider the hyperelliptic algebraic curve $\L$ of genus $g$ defined  by
\be
w^2 =(\l-\x)(\l-\xb)\prod_{j=1}^{g} (\l-E_j)(\l-F_j)
\la{L}\ee
with $\x=x+i\rho$
symmetric with respect to antiholomorphic involution $\l\ra\bar{\l}$ that
entails for some $m\leq g$
\ben
E_j=\bar{F_j}\;\;,\;\; j=1,\dots, m 
\een    
\be
E_j, F_j\in \R\;\;,\;\; j=m+1,\dots, g
\la{real}\ee
Introduce on $\L$ the canonical basis of cycles 
$(a_j,\;b_j)\;\;\; j=1,\dots g$. 
Each cycle $a_j$ is chosen to surround the branch cut $[E_j,\;F_j]$;
cycle $b_j$ starts on one bank of the branch cut $[\x,\xb]$, goes on the
other sheet through branch cut $[E_j,\;F_j]$ and comes back.
The dual basis of holomorphic differentials
$d U_j\;,\;j=1,\dots, g$ is normalized by
\be
\oint_{a_j} dU_k =\delta_{jk}
\la{norm}\ee
Define $g\times g$ matrix of b-periods 
${\bf B}_{jk}=\oint_{b_j}dU_k$ and related
$g$-dimensional theta-function $\T(z|{\bf B})$.
Differentials $dU_j$ are linear combinations of non-normalized 
holomorphic differentials
\be
dU^0_j = \f{\l^{j-1} d\l}{w}\;\;\;\; j=1,\dots, g
\la{nonorm}\ee
General algebro-geometrical solution of the Ernst equation 
may be written in many different forms (see \c{K,K1}). Here it is
convenient to use the original form of \c{K}:
\baa
\E=\f{\T(U|_D^{\i^2} + B_\O - K)\T(U|_D^{\i^1}-K)}
{\T(U|_D^{\i^1} + B_\O - K)\T(U|_D^{\i^2}-K)}
\nn\\
\times \exp\{\O|_{\i^1}^{\i^2}\}
\la{SOL}\eaa
where the new objects are defined as follows: 
$K$ is a vector of the Riemann constants of $\L$; 
$D$ is a set (divisor)
of $g$ $(\x,\xb)$-independent  points $D_1,\dots, D_g$ on $\L$;
\ben
\left(U|_D^{\i^{1,2}}\right)_k\equiv \int_{P_0}^{\i^{1,2}}
dU_k -\sum_{j=1}^g \int_{P_0}^{D_j} dU_k
\een 
with an arbitrary base point $P_0$ (entering also vector $K$).

It remains to define $\O(\i^1)-\O(\i^2)$ and vector $B_{\O}$.
Let $d\O(P)$ be an arbitrary locally holomorphic
1-form on $\L$ with $(\x,\xb)$-independent singularities and
related singular parts satisfying the 
normalization conditions
\be
\oint_{a_j} d\O =0\;\;\;\;\; j=1,\dots, g
\la{normO}\ee
Define its vector of $b$-periods 
\be
(B_\O)_j =\oint_{b_j} d\O
\la{bO}\ee
and require the reality conditions
\ben
\O(\i^2)-\O(\i^1)\in\R\;\;\;\;\;\;\; {\rm Re} (B_\O)_k =\pm \f{1}{4}
\een
Now solution \Ref{SOL} is completely defined. If one take 
$g=0$ then combination of theta-functions in \Ref{SOL} disappears and
we get 
\be
\E=\exp\left\{\O_0 (\i^2)-\O_0(\i^1)\right\}\in \R
\la{staticsol}\ee
 i.e. static solution,
which serves as a static background of solution \Ref{SOL}.
 It is easy to show that
by an appropriate choice of differential $d\O_0$ on
the Riemann surface $\L_0$ given by 
\ben
w^2 = (\l-\x)(\l-\xb)
\een
one can get arbitrary static
solution. Namely, take an arbitrary  
solution $\E_0\in\R$ (for definiteness, asymptotically flat 
i.e. $\E_0(\x=\i)=1$) satisfying
\Ref{static} and define  1-form $d\O_0$ by
\ben
d\O_0 (\l,\x,\xb)
\een
\be
=\f{d\l}{4}\int_{\i}^\x d\x'
\f{\p}{\p\l}\left(\sqrt{\f{\l-\xb'}{\l-\x'}}\right)
\f{\p\log\E_0(\x',\xb'}{\p\x'}
\la{def1}\ee
and analogous equation with respect to 
$\xb$ (which are compatible as a corollary of
\Ref{static}); we have 
\ben
d\O_0 (\l) = \overline{d\O_0(\bar{\l})}
\een
This is a simple example of "direct scattering procedure"
(and  analog of Fourier transform):
the positions and structure of singularities of $d\O_0$ carry the
whole information about solution $\E_0$. 

The  1-form $d\O$ on $\L$ which enters \Ref{SOL} inherits all 
singularities of $d\O_0$ on $\L_0$ and is assumed to have
additional simple poles at the branch points $E_j$ with the residues $1/2$
and at the branch points $F_j$ with the residues $-1/2$, $j=1,\dots, g$.

Therefore, for fixed genus $g$ 
the solution \Ref{SOL} is defined by the following set of data:
an arbitrary background solution 
$\E_0$ of the Ehlers-Darboux equation \Ref{static}
and $(\x,\xb)$-independent points
$\{E_j\;, F_j\;, D_j\;\;\; j=1,\dots, g\}$.

\section{Reduction to Meinel-Neugebauer construction}

To obtain the solutions constructed 
in \c{MN,MN1} as a special case
 of \Ref{SOL} one have to take $m=g$ i.e. for all 
$j=1,\dots,g$ one  assume
\be
F_j = \bar{E_j}
\la{vert}\ee
Then to rewrite solutions \Ref{SOL} in the form of \c{MN}
introduce on $\L$  meromorphic 1-form $dW$ having the 1st order
poles at $\l=\i^1$ and $\l=\i^2$ with the residues $-1$ and $+1$
respectively normalized by
\ben
\oint_{a_j} dW=0\;\;\;\;\; j=1,\dots, g
\een
The following simple identity:
\ben
\exp\{W(\Dt)-W(D)\}
\een
\ben
\;\;\;\;\;\;\;\equiv\f{\T(U(\i^2)-U(\Dt) - K)}
{\T(U(\i^1)-U(\Dt) - K)}
\een
\be
\;\;\;\;\;\;\;\;\;\;\;\;\;\;\;\;\;\; \times
\f{\T(U(\i^1)-U(D)-K)}
{\T(U(\i^2)-U(D)-K)}
\la{id}\ee
is valid for arbitrary two sets of $g$ points $D$ and $\Dt$ and may be
verified by simple comparison of the pole structure of both sides
with respect to every $D_j$ and every $\Dt_j$.

Thus solution \Ref{SOL} may be rewritten as follows:
\be
\E=\exp\left\{W|_{D}^{\Dt} + \O|_{\i^1}^{\i^2}\right\}
\la{SOL1}\ee
where divisor $\Dt$ consists of the points $\Dt_1,\dots,\Dt_g$ defined
by the following system of equations:
\be
U(\Dt)-U(D)=-B_\O
\la{JP}\ee
The vector in the l.h.s. is understood as
\ben
(U(\Dt)-U(D))_k=\sum_{j=1}^g \int_{D_j}^{\Dt_j} dU_k
\een
The problem of determining points of $\Dt$ from 
 \Ref{JP} is called the Jacobi inversion problem.

Equations \Ref{JP} may be rewritten in terms of non-normalized basis of
holomorphic differentials given by \Ref{nonorm} 
as follows:
\be
\sum_{k=1}^g\int_{D_k}^{\Dt_k} \f{\l^{j-1}d\l}{w} =
\oint_{\p\Lh}\O \f{\l^{j-1}d\l}{w}
\la{JP1}\ee
for $j=1,\dots g$
where $\p\Lh$ is the boundary of $4g$-sided fundamental polygon $\Lh$ of
surface $\L$ which is obtained if we cut $\L$ along all basic cycles;
\ben
\O(P)\equiv\int_{P_0}^P d\O\;\;\;\;\; P\in\L
\een
with arbitrary base point $P_0\in\L$; choice of $P_0$ does not influence the
r.h.s. of \Ref{JP1}.
Expression \Ref{JP1} may be easily derived from \Ref{JP} using the
general formula valid for any two 1-forms $W_{1,2}$ on $\L$ \c{Mum}:
\be
\oint_{\p\Lh} W_1 dW_2 = \sum_{j=1}^g \left\{A_{W_1}^j B_{W_2}^j -
B_{W_1}^j A_{W_2}^j\right\}
\la{formula}\ee
where $A_{W_{1,2}}^j$ and  $B_{W_{1,2}}^j$ are $a$ and $b$ periods of the
forms $dW_{1,2}$ (to derive 
\Ref{JP1} one should take $dW_1=d\O$, $dW_2=dU_j^0$).

Introducing differential 
\ben
dW^0\equiv
\f{\l^g d\l}{w}
\een
(which coincides with $d W$ up to some combination of holomorphic
differentials \Ref{nonorm} which provide vanishing of all $a$-periods
of $dW$), and applying \Ref{formula} to $d\O$ and $d W_0$, we rewrite
\Ref{SOL1} as follows:
\be
\E=\exp\left\{W_0|_D^{\Dt}
+\oint_{\p\Lh}\O dW_0 + \int_{\i^1}^{\i^2} d\O\right\}
\la{SOL2}\ee
Formulas \Ref{JP1} and \Ref{SOL2} 
after identification
\be
u_j\equiv  \oint_{\p\Lh}\O \f{\l^{j}d\l}{w} \;\;\;\;\;\;\; j= 0 ,\dots, j-1
\la{MN1}
\ee
\be
u_g\equiv \oint_{\p\Lh}\O \f{\l^g d\l}{w} + \int_{\i^1}^{\i^2} d\O
\la{MN2}
\ee
may be rewritten in the following way:
\be
\sum_{k=1}^g\int_{D_k}^{\Dt_k} \f{\l^{j-1}d\l}{w} = u_j\;\;\;\;\;\;
j=0,\dots,j-1
\la{JP2}\ee
and
\be
\E=\exp\left\{\sum_{j=1}^g \int_{D_j}^{\Dt_j}\f{\l^g d\l}{w} + u_g\right\}
\la{SOL3}\ee
which precisely coincide with expressions of \c{MN}. Functions 
$u_j\;,\; j=1,\dots, g$ satisfy the Laplace equation
\ben
\Delta u_j =o
\een
and the recurrent equations
\be
u_{j\;\x}=\f{1}{2}u_{j-1} +\x u_{j-1\;\x}
\la{MNeq}\ee
as a corollary of the relations
\ben
\Delta \f{1}{w} =0
\een
\ben
\left(\f{\l^j}{w}\right)_\x=\f{1}{2}\f{\l^{j-1}}{w}  +
\x\left(\f{\l^{j-1}}{w}\right)_\x
\een
and the residue theorem applied to the contour integral over $\p\Lh$.

The static background of solution \Ref{SOL3} is given by an arbitrary
solution of the Laplace equation $u_g$ (one could take any other 
function $u_j$, since it would almost uniquely determine the others
according to \Ref{MNeq}), which may, therefore,
be alternatively expressed as
\ben
u_g= \int_{\i^1}^{\i^2} d\O_0
\een
in terms of the differential $d\O_0$ \Ref{def1}.

Let us show how to choose the parameters of the 
present construction to get the ``dust disc" solution of \c{MN1} 
posed at $(x=0\;,\;\rho\leq \rho_0)$. 

One should take $g=2$, choose some complex
$E_1$ (related to parameter $\mu$ of \c{MN1}) and put  $F_1=\bar{E_1}$;
$E_2= -F_1$; $F_2=\bar{E_2}$.
The 1-form $d\O$ should be taken in the form
\ben
d\O(\l)= d\hat{\O}+d\tilde{\O}
\een
where
\be
d\hat{\O}\equiv
\int_{-i\rho_0}^{i\rho_0} f(\g)d\O^{(\g)}(\l)) d\g
\la{OH}
\ee
with the integral  taken along the imaginary axis;
$d\O^{(\g)}(\l)$ is meromorphic 1-form on $\L$ with vanishing $a$-
periods and unique pole of the second order at $\l=\g$ with leading
coefficient equal to 1;
$f(\g)$ may be an arbitrary measure satisfying $\bar{f}(\g)= f(\bar{\g})$
(for example, $f\in\R$). $d\tilde{\O}$ is a meromorphic 1-form on $\L$
having simple poles with the residues $-1$ at $E_{1,2}$ 
and $+1$ at $D_{1,2}$.
 Related static background will be given by \c{K}
\ben
\log \E_0= \int_{-i\rho_0}^{i\rho_0} \f{f(\g)d\g}{\{(\g-\x)(\g-\xb)\}^{1/2}}
\een
Specifying $f(\g)$ in some special way (see \c{MN1}) 
one arrives to the ``dust disc" solution of \c{MN1}.

Formula \Ref{SOL} for the Ernst potential may now be rewritten as follows
(see \c{geom}):
\be
\E=\f{\Theta(U|_{\xi}^{\i^2} +B_{\hat{\O}})}
{\Theta(U|_{\xi}^{\i^1} +B_{\hat{\O}})}\exp\hat{\O}|_{\i^1}^{\i^2}
\la{SOL10}
\ee
where $2\pi i B_{\hat{\O}}$ is the vector of b-periods of $d\hat{\O}$.

\section{Summary}
We have shown that the solutions 
of the Ernst equation 
obtained recently in \c{MN} (and, in particular,
 some partial solution of this class
exploited in \c{MN1} to describe rigidly rotating dust disc)
constitute a subclass of the 
algebro-geometric (finite-gap) solutions found before in \c{K}.

In spite of the solutions derived in \c{MN,MN1} are not new,
the physical interpretation of special solution of
this class proposed in \c{MN1} would be very interesting if it would really
describe the dust disc. However, the rings $\xi=E_j$ are most probable
the singular points of the Weyl scalars, even though the metric
coefficients may be finite and differentiable once at these rings.
One of the main reasons to expect the singularity of these rings is that 
the generic asymptotical expansion of $\E$ at $\x=E_j$ contains 
logarithmic singular terms (this is a standard hypergeometric-like
asymptotical expansion near the singularity). By the same reason
it is not appropriate for numerical simulation at these points. 
The only way to prove non-singularity 
or singularity of solution of \c{MN1} on the rings would be to find 
explicit asymptotical expansion 
of related Weyl scalars at $\xi=E_j$. Before this is done
the physical interpretation of solution \Ref{SOL10},\Ref{OH} as the field of 
rotating dust disc will remain questionable.

Concluding, we  express the hope that 
some of the finite-gap solutions will find reasonable physical
application (see also \c{K1} for discussion, where, in particular,
we describe a solution with toroidal ergosphere).

{\bf Acknowledgments} I thank R.Meinel for sending me copy of \c{MN}
and discussions.

\end{document}